\documentclass[aps,prl,twocolumn,groupedaddress]{revtex4}
\usepackage{graphicx}

\def\ll{\label}
\def\re{\ref}
\def\c{\cite}

\def\r1{(\ref{$1})}

\def\sn{\rm sn}

\def\cn{\rm cn}
\def\dn{\rm dn}

\def\th{\theta}
\def\ba{\begin{array}{c}}

\def\ea{\end{array}}

\def\si{\sigma}

\def\bet{\beta}
\def\ov{\over}
\def\ha{{1\over 2}}

\def\l{\left}
\def\l({\left(}
\def\r){\right)}
\def\r{\right}

\def\la{\lambda}
\def\al{\alpha}

 \def\be{\begin{equation}}
\def\bc{\begin{center}}
\def\ec{\end{center}}
\def\bit{\begin{itemize}}
\def\eit{\end{itemize}}
\def\ee{\end{equation}}
\def\ed{\end{document}}
\def\bea{\begin{eqnarray}}
\def\eea{\end{eqnarray}}
\def\efr{\end{flushright}}

\begin{document}
\title{Integrable  multi atom matter-radiation models  without
rotating wave approximation}

\author{
Anjan Kundu 
}\affiliation{
  Saha Institute of Nuclear Physics,  
 Theory Group \\
 1/AF Bidhan Nagar, Calcutta 700 064, India.}
 \email
{email: anjan@tnp.saha.ernet.in} 


%
\begin{abstract} 
Interacting   matter-radiation models 
 close to  physical systems are proposed, which without
 rotating wave approximation  and with  matter-matter 
 interactions are Bethe ansatz  solvable.
 This integrable system is  constructed from the elliptic
 Gaudin model at  high spin limit, where 
 radiative excitation  can be included   
perturbatively. 


\end{abstract}
\maketitle

In dealing with interacting matter-radiation systems the counter rotating
wave (CRW) terms, which inevitably appear in  general situation, are
usually neglected invoking the rotating wave (RW) approximation,
  without which physical models  generally become unsolvable.
 However  the RW  approximation
breaks down away from the resonance condition at  $\omega_a\approx 
\omega_f $ as well as  for high  intensity  fields.
 Consequently,
 fast oscillations (with frequency $\omega_a+\omega_f $) 
 can no longer  be neglected compared to the slow ones (with frequency
$|\omega_a-\omega_f |$) and
 one is forced to consider the general case with CRW terms. Moreover, since
the situation is generic, this problem can arise in diverse models like
those in quantum optics,
 in cavity QED both in microwave and optical domain
\c{rempe8790,raizen89},  
in trapped ions irradiated by  laser beams
 \c{trap}
as well as in transport through quantum
dots coupled to boson model \c{qdot03}
and  in quantum information transfer protocol with a superconducting circuit
\c{qinf03}.

Under RW approximation one can obtain 
 exactly  solvable models like Jayenes-Cummings (JC) \c{jc},   
Buck-Sukumar \c{bs} model and their multi-atom generalizations
\c{njc,nbs,kundu04}  as well as 
 $q$-deformed matter-radiation models, inducing anisotropic  
 together  with higher nonlinearity
\c{qjc,kundu04}. 
However when this approximation is avoided,  CRW terms appear 
having
  the form like 
$ \ H_{crw}= 
 \bet (b^\dag \si^++ b \si^-)  
 \ $, in the simplest case
and    generally spoil  the  solvablity of the system.
Models with  CRW terms in various forms  were 
investigated  earlier  \c{trap,crw03,prl83}, though  such exactly 
solvable multi-atom models which are  close to physical systems
are not known in the literature,  except perhaps a   
 recent proposal \c{amico} involving some unphysical 
  terms.

We propose here
   integrable  multi-atom JC type matter-radiation models which include
CRW terms  and allow exact   Bethe ansatz solutions 
for the vanishing radiation frequency, when 
    the RW  approximation is naturally  not applicable. Field excitation term
   can  be considered  over the integrable model, perturbatively 
taking  $\omega_f$ to be small, while the interatomic
interactions through spin-spin coupling can be included in an exact way.
We derive our
 model
 from the elliptic Gaudin model \c{egaudin}, 
through  spin-$\ha$ representation
for  the atoms   and the bosonic realization   under 
   high spin limit for the
single mode radiation field.
For following the logic of the construction, we recall that all 
known exactly
solvable  multi-atom  matter-radiation 
models {\it with}
 RW approximation are derived
  from integrable $xxx$,  $xxz$ spin
models \c{njc,nbs,kundu04}  or from their corresponding Gaudin limits 
\c{jurco,duk04}.
 Note that at  higher spins the underlying algebras associated 
with  these  models are  either $su(2), su(1,1)$
 or their quantum deformations, both  
 allowing the crucial bosonic realization.
 For possible construction of 
integrable models {\it without}
 RW approximation, one may  therefore expect to repeat 
the same construction 
starting from a more general integrable  
inhomogeneous $xyz$ model with higher spin 
representation \c{takabe,xyz}.
 The representative Lax operator of  this model depends,
 apart from  the spectral
parameter $\la $ and the  anisotropy
parameter $\alpha$, on  the elliptic modulus $k$,
 inhomogeneity parameters $z_n$  as well as on the 
spin $s$ representation, through its dependence on
  coefficients
  $W_p (\la-z_n;\alpha,k), \ p=0,1,2,3 $, 
  expressed through elliptic functions and  operators
 $S^p(s,\alpha,k)$ satisfying the Sklyanin algebra \c{sklyalg}.
In fact, all known integrable matter-radiation 
models mentioned above can be obtained
from this general structure at  different limits of the parameters
involved.  
For example, at $k=0$,
  one recovers the  trigonometric 
$xxz$ case, when   generators of the Sklyanin algebra 
  reduce to 
those of the quantum algebra: $S^a(s,\alpha), a=1,2,3$.
Under a further limit of $\alpha \to 0$ one obtains
the trigonometric Gaudin model
   with the operators reducing  
to  $S^a(s)$, satisfying the standard  Lie algebra. If however
at the same time
a  limit $\la \to 0$ 
 is   imposed   on the spectral parameter (together
with $z_n$), 
 we recover the  
$xxx$ case as well as 
the corresponding rational Gaudin model. 

We   however
  keep the elliptic modulus $k$
nontrivial  throughout our construction and
   thereby follow a  rout different  from the earlier ones.  
Observe  that,  unlike the above cases
the Sklyanin algebra
  does not have any known
 bosonic realization, though  fortunately
  at $\alpha \to 0$ it reduces
   to the standard algebra 
allowing  bosonic mapping through
 the  Holstein-Primakoff
transformation (HPT):
$S^+= b^\dag \sqrt{ {s}-\rho \hat N }, \
S^-= \sqrt{ {s }- \rho \hat N} b  , \ S^3= \hat N- \rho {s \ov 2}, \
 \hat N=b^\dag b$,
where $\rho =\pm$ correspond to $su(2)$ ($su(1,1)$).
 At this limit the $xyz$ model reduces
to  the integrable  elliptic Gaudin model \c{egaudin}
 with mutually commuting  conserved quantities 
\be H_n=\sum_{a,m \neq n} w_a(nm) S^a_nS^a_m,
\ll{hegaudin}\ee 
{with} $ \ w_1(\la)={ \cn \ov \sn}(\la), \ 
 w_2(\la)={ \dn \ov \sn}(\la), \
 w_3(\la)={ 1 \ov \sn}(\la)
\ $
and  $\sum_n H_n=0$, where we have introduced  short hand notation 
 $w_{a}(nm) \equiv w_a(z_n-z_m) $.
However, we   face now the difficulty, that  for 
 $\rho =+ $ the 
 operators  $S^\pm$   in  the HPT  become
nonhermitian  at $ <\hat  N> \ \ > {s }$, while for
 $\rho =- $, though this is  avoided, Gaudin
 Hamiltonian (\re{hegaudin}) becomes nonhermitian to maintain
   its integrability. We resolve this problem
 by taking the spin limit
 $s_0 \to \infty $   in HPT for  the radiation field at $n=0$,
yielding 
\be 
S^+_0={ {1 \ov   \epsilon }} b^\dag, \
S^-_0=  { {1 \ov   \epsilon }} b  , \ S^3_0=- {1 \ov  2\epsilon^2}, \quad 
\epsilon = {1 \ov  \sqrt s_0} \to 0.
 \ll{hpt1}\ee
 We have 
considered here   $\rho=+1 $  for definiteness and kept terms 
up to order
$O({1\ov \epsilon})$. For modeling  $N_a$ 
two-level atoms
we  take spin $s_j=\ha$ representation at all other points
 $ j=1,2, \ldots,N_a$.
The  ${1\ov \epsilon}$ coefficients in the expansion  of the elliptic Gaudin 
Hamiltonians, though have the desired 
 CRW terms,  not yet yield  the mutually commuting conserved set.  
 The reason for this is twofold: 
 the appearance of  $xyz$ type spin terms  $\si^+_j\si^+_k+ cc.$
 with  coefficients $w_-(jk)= \ha (w_1(jk)- w_2(jk))$
and   the inhomogeneity  $w_3(j0)$ in the  $\si ^3_j$ term, both of  which
  we have to tackle before extracting 
integrable models  from such an  expansion.
 Obviously the first  difficulty  can be
   trivially resolved by setting $k = 0$, which  yields  
  $w_-(jk) = 0 $ as considered in \c{jurco,duk04}. However, since
 our aim  is
to keep  the elliptic modulus $k$  nontrivial, we have to take a different 
rout.   Our
  strategy    would be to push the undesired $w_-(jk)$ term out 
from the given order by 
  suitably  scaling  the
inhomogeneity  parameters as
$ \ z_0=K+\epsilon x_0, \ z_j=\epsilon x_j, \ j=1,\ldots N_a  
\ $
 and taking
 the limit $\epsilon  \to 0 $,
 $K$ being the elliptic integral \c{xyz}. 
Observing that $su(2)$   is isomorphic  under 
the reflection  of any axis  of the  basis vectors, we also   redefine 
the coupling constants in (\re{hegaudin}) 
as  $ \ w_2(jk) \rightarrow -w_2(jk), \ \
 w_1(jk) \leftrightarrow w_3(jk), \ $ and similarly for $w_a(0j)$
, to have 
 more conventional notation.  
The redefined coupling constants using 
 $w_a(0j)=-w_a(j0)$ and the property  of the elliptic functions,  reduce  
 in  the needed order to 
\begin{widetext}\be
w_1(0j) \to 1, \   w_2(0j) \to 
{k^{'}}, \ w_3(0j) \to 
0 , \ \ w_a(jk) 
 \to 
 {1\ov \epsilon (x_j-x_k)}, \ a=1,2,3
, \ \mbox{with} \ \  k^{'}=(1-k^2)^\ha  .
 \ll{w}\ee\end{widetext}
 This shows  that in the given order 
   now we have $w_-(jk)=0, w_3(j0)=0$, which  
 simultaneously removes   both the 
 above obstacles and  derives  finally from (\re{hegaudin})
in the first nontrivial order $O({1 \ov  \epsilon})$,
 the  integrable  matter-radiation models with 
 mutually commuting  Hamiltonians  $ [H_j,H_k]=0$: 
\be 
   H_j = \omega_a \si_j^3+ H_j^{b\si}+ H_j^{\si \si}, 
\   
\ll{hjc}\ee  with $j=1, \ldots, N_a $, where \be
H_j^{b\si}= \Omega_+ (b\si^+_j+b^\dag
 \si ^-_j)+ \Omega_- (b^\dag \si^+_j+b \si^-_j),  
\ll{hjcb}\ee
with $ \Omega_\pm=  1\pm k^{'}$, describes 
 interaction between atoms and the radiation    
 with explicit CRW terms and
\be H_j^{\si \si}=
\sum^{N_a}_{k \neq j} {1\ov x_j-x_k}( \si ^+_j\si ^-_k+
 \si ^-_j\si ^+_k+ \si ^3_j\si ^3_k),
\ll{hjcs}\ee
models interatomic interactions, but without having any
  $xyz$ type 
spin term. This crucial  fact permits us 
to   include 
the atomic inversion term with arbitrary coefficient $  \omega_a $
in (\re{hjc}), since $ \si_j^3$
 commutes  with the whole Hamiltonian.
 The coupling constant $\Omega_-$ for the interaction with CRW terms  
   clearly vanishes at   
 $k=0$,     
recovering   the earlier results  with RW approximation 
 \c{jurco,duk04}.
Note that, we can construct a series of integrable Hamiltonians
by different combinations of the commuting set $H_j$.
  For example, a generalization of 
the Tavis-Cummings model \c{njc}  without   RW approximation
can be constructed  as
\bea
H_0=  \sum_jH_j &=&\sum_j^{N_a}[ \omega_a\si^3_j+ 
\Omega_+ (b\si^+_j+b^\dag
 \si ^-_j)   \nonumber \\ 
&+& \Omega_- (b^\dag \si^+_j+b \si^-_j)], 
\ll{gtc}\eea
without   having explicit       
interactions between the atoms.
 At $k=1$, when $\Omega_+=\Omega_-=1$,
(\re{gtc})   reduces further to 
\be
H_1= \sum_j^{N_a}( \omega_a\si^3_j+ 
(b+b^\dag) (\si^+_j+ \si^-_j )), \
\ll{gtc1}\ee
which appears  in the trapped ion model interacting with the
 center  of mass motion,  after taking the customary 
 Lamb-Dicke limit.
Notice that (\re{gtc1})
 is similar to  the model studied in \c{crw03}, 
if a bosonic number operator term 
is added to it. However,
 while  the model  in   \c{crw03}   becomes integrable 
only  at the thermodynamic and the mean field limit of the atoms, 
our model (\re{gtc1})  achieves this without going to such limits.  
Moreover, contrary to the popular belief, that 
such a model with finite  $N_a$  atoms is exactly
  solvable  only under RW
approximation  and otherwise chaotic \c{prl83}, we show it to be 
 Bethe ansatz solvable in general, at least for $\omega_f=0$.

Since  our model is derived from the elliptic Gaudin model  through
some limiting procedure, we can obtain its  
 Bethe ansatz formulation also 
from the related result \c{egaudin}, by   taking suitably 
the limits
of the parameters involved.
One  important consequence of this inheritance is that, the excited Bethe eigenstates $|M>$
, as in
 the elliptic Gaudin model, are no
 longer 
arbitrary, but constrained by the total spin value $M=\sum_n s_n$.
 For our model therefore, we must have: $M =\ha N_a+s_0 \to {1/\epsilon ^2}$
 due to the limit
$s_0={1/\epsilon ^2} \to \infty$   for the bosonic mode.
 This macroscopic excitation with high photon number is in conformity with
 the results of \c{crw03}, indicating that we must be
 in the super-radiant phase with no possibility of phase transition, since 
 due to $\omega_f=0$ the critical coupling here would be
$\sqrt{\omega_f \omega_a} =0$.
Inspite of the fact that, the
  investigation  of \c{crw03} is valid only for 
nontrivial $\omega_f$ and its approach is
totally  different  from ours, 
some apparent similarity between    these results in the integrable case,
 perhaps  may  be explained by the point that  
of the thermodynamic limit adopted   
 in \c{crw03} is mimicked by the high spin
limit for the bosonic mode,  considered in our model.

For constructing  exact eigenstates and eigenvalues, 
we  introduce     scaling also for  the two sets of
   Bethe parameters: $w_b =\epsilon l_b, b=1, ...{N_a \ov 2}$ and
$w^0_\alpha =\epsilon l^0_\alpha+ K_, \al=1,\ldots, {s_0},$  
similar to those for the inhomogeneity parameters defined above.
 The    
 exact eigenvalue  
for our matter-radiation models (\re{hjc}) is then   derived from the
 limiting values of the Bethe ansatz result
of \c{egaudin}  as
\begin{widetext}
\be E^{(crw)}_j=  \th _{11}^{'}(0) \left(
 \omega_a+\sum_{b=1}^{N_a \ov 2} { 1 \ov  (x_j-l_b) }
+\ha \sum_{k \neq j}^{N_a} { 1 \ov  (x_j-x_k)}+
{\th _{10}^{''}\ov \th _{10}} (0)(
2x_j-x_0 -S )\right), \ 
\ll{evcrw}\ee
\end{widetext}
where $S=\epsilon^2\sum^{s_0}_\alpha l^0_\alpha $.
Possible   nontrivial contribution  in the given order   may be
obtained for $S$ as $S= l^0$,  
when  $ l^0_\alpha= l^0$ 
 are degenerate for all $ \alpha=1,\ldots, {s_0} $, which
 we  consider here for definiteness.
 From the 
  Bethe equations for $ l^0_\alpha $
 we find a  solution for these
degenerate parameters as $l^0=x_0$, while for the rest of
 the Bethe parameters $l_b, \quad b=1,\ldots, {N_a \ov 2},$ one needs to 
solve the equations
\be 
 \ha \sum_{k=1}^{N_a} 
 {1  
\ov  (x_k-l_b) }=\omega_a+
 \sum_{c \neq b}^{N_a \ov 2}
 { 1 
\ov  (l_c-l_b) }, 
\ll{becrw}\ee
for a given set of arbitrary  
   inhomogeneity parameters  $x_j, j=1,\ldots,N_a$.
For practical relevance we have  applied
this   Bethe ansatz result to a $N_a=2$ ion model
  with CRW terms 
described by (\re{hjc}) and shown its exact energy spectrum 
for $j=1$  in  Fig. 1. Similar result
can be obtained easily for $j=2$, by interchanging the parameters $x_1
\leftrightarrow x_2$.  
 \begin{figure}
\includegraphics[width=7.5cm]{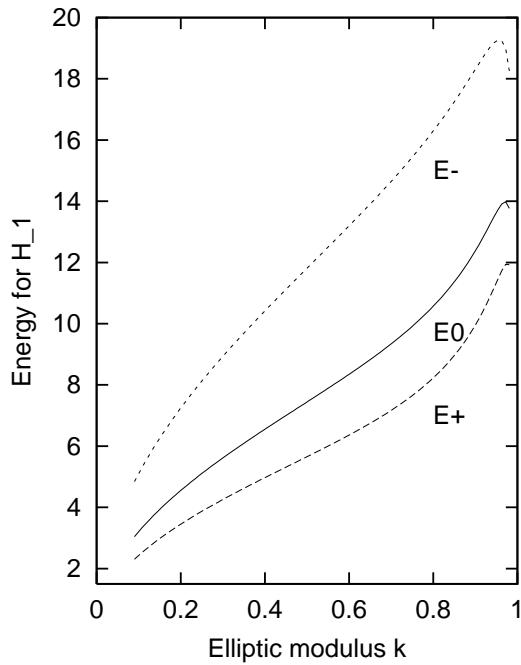}
 \caption{\label{}  Vacuum $E0$ and excitation energies $E\pm$
showing asymmetric Rabi-type energy splitting, 
for different values of  elliptic modulus $k \in [0,1]$.
  The graphs depends crucially on the 
inhomogeneity parameters (chosen here as $x_0=1, x_1=1.1, x_2=3$)
and the atomic frequency (taken  as $\omega_a=1$). However
the qualitative behavior appears to remain the same:
all the energies rapidly increasing   with the increase of
 $k$,   with a curious dip toward the end.}
 \end{figure}

 For a  
 closer contact with    physical systems we may include  
 a  radiative excitation term $\omega_f \hat N$,   perturbatively, 
over our   Bethe ansatz solvable  models (\re{hjc},\re{gtc},\re{gtc1}).
For  smaller  values of  field frequency 
$\omega_f $, when  the RW approximation  worsens, the
 perturbative treatment would become more accurate.
 One can  generalize the formulation of  standard
  perturbation theory for applying 
it to   integrable models with exact eigenvalues 
 $E_n(\vec l )$ like (\re{evcrw}) and the corresponding
 normalized Bethe eigenstates
$\psi(\vec l) $, with Bethe parameters $\vec l \equiv (l^0_1, \ldots, 
l^0_{s_0},l_1,\ldots,  l_{{N_a \ov 2}}) $. 
 The result
would however depend on the concrete model and might be  difficult to
extract
in the explicit form.  For  considering  the bosonic number operator 
 term
we are interested in,   denoting $  N (\vec l,\vec m ) =
<\psi(\vec l) \hat N \psi(\vec m)>$,
we can derive 
 the first order perturbative  correction 
to the exact eigenvalues as    
 $\omega_f   N (\vec l,\vec l^{} )$
 and to the  corresponding  eigenstates as
$\omega_f \sum_{\{\vec m \}}  
{N(\vec{ m}, \vec {l}) \ov E_n(\vec l)-
E_n(\vec {m} )}  \psi(\vec m), \ $ where
the  sum is over
the whole solution set of the Bethe parameters.


Thus   we have  constructed  and exactly solved
multi-atom matter-radiation models without rotating wave approximation
and with explicit interatomic interactions. We 
derive our integrable models and the corresponding
exact Bethe ansatz result
  from the 
elliptic Gaudin model through a limiting procedure.
 We can include  
physically important  field excitation
 term perturbatively, over the exactly solvable
models. 

 \end{document}